\documentclass[prd,nofootinbib,english]{revtex4-1}
\usepackage{amsmath}
\usepackage{multirow}
\usepackage{amsfonts,color}
\usepackage{amssymb,float}
\usepackage[utf8]{inputenc}
\usepackage{graphicx}
\usepackage{babel,blindtext}
\usepackage{color}
\usepackage[unicode=true,pdfusetitle,
 bookmarks=true,bookmarksnumbered=true,bookmarksopen=true,bookmarksopenlevel=3,
 breaklinks=false,pdfborder={0 0 1},backref=false,colorlinks=true]
 {hyperref}
\usepackage{enumitem}
\usepackage{subfigure}

\usepackage{color}

\begin{document}

\title{Dynamical system analysis of generalized energy-momentum-squared gravity}


\author{Sebastian Bahamonde}
\email{sbahamonde@ut.ee, sebastian.beltran.14@ucl.ac.uk}
\affiliation{Laboratory of Theoretical Physics, Institute of Physics, University of Tartu, W. Ostwaldi 1, 50411 Tartu, Estonia}
\affiliation{Department of Mathematics, University College London, Gower Street, London, WC1E 6BT, United Kingdom}

\author{Mihai Marciu}
\email{mihai.marciu@drd.unibuc.ro}
\affiliation{Faculty of Physics, University of Bucharest, 405 Atomistilor, POB MG-11, RO-077125 Bucharest-Magurele, Roumania}

\author{Prabir Rudra}
\email{prudra.math@gmail.com, rudra@associates.iucaa.in}
\affiliation{Department of Mathematics, Asutosh College, Kolkata-700026, India.}



\date{\today}

\begin{abstract}
In this work we have investigated the dynamics of a recent modification to the general theory of relativity, the energy-momentum squared gravity model $f(R,\mathbf{T^2})$, where $R$ represents the scalar curvature and $\mathbf{T^2}$ the square of the energy-momentum tensor. By using dynamical system analysis for various types of gravity functions $f(R,\mathbf{T^2})$, we have studied the structure of the phase space and the physical implications of the energy-momentum squared coupling. In the first case of functional where $f(R,\mathbf{T^2})=f_0 R^n(\mathbf{T^2})^m$, with $f_0$ constant, we have shown that the phase space structure has a reduced complexity, with a high sensitivity to the values of the $m$ and $n$ parameters. Depending on the values of the $m$ and $n$ parameters, the model exhibits various cosmological epochs, corresponding to matter eras, solutions associated with an accelerated expansion, or decelerated periods. The second model studied corresponds to the  $f(R,\mathbf{T^2})=\alpha R^n+\beta\mathbf{(T^2)}^m$ form with $\alpha, \beta$ constant parameters. In this case, a richer phase space structure is obtained which can recover different cosmological scenarios, associated to matter eras, de--Sitter solutions, and dark energy epochs. Hence, this model represents an interesting cosmological model which can explain the current evolution of the Universe and the emergence of the accelerated expansion as a geometrical consequence.  

\end{abstract}


\maketitle

\section{Introduction}
The discovery of the accelerated expansion of the Universe \cite{Perlmutter:1998np, Spergel:2003cb} at the turn of the last century has come as the most unexpected and surprising result for the scientific society. The reason behind the reaction is quite obvious as gravity being an attractive force will compel the Universe and all the matter present inside it to contract. So the expansion of the Universe would gradually slow down and finally reach a situation where it stops totally. After this gravity will pull it back and the Universe would undergo a contraction. But the observational evidences are saying a totally different story. This is where our standard knowledge of physics is falling short and we are compelled to search for some new physics which will help us to explain the phenomenon that our observations are showing. 

Over the last two decades, the scientific society have left no stones un-turned in its quest for a suitable physical theory that will explain the accelerated expansion~\cite{Copeland:2006wr} of the Universe. The whole effort can be broadly classified into two categories. The first one targets the nature of matter that fills the Universe. This theory tells us that the Universe is filled with a mysterious negative pressure component termed as ``dark energy" which provides an antigravitating stress to sustain not only an expanding Universe but also fuels it to reach a level of accelerated expansion. The most common way of doing this is by introducing a cosmological constant $\Lambda$ in the Einstein's equations of general relativity. The idea seems to be consistent but the concept is still in its infancy and is plagued by a lot of shortcomings. The most prominent one being its invisible nature and hence the term ``mysterious". Moreover there is the cosmological constant problem to deal with \cite{Martin:2012bt} which is associated to the lack of consistency between the observed values of vacuum energy density (given by a very small value of the cosmological constant) and the theoretically predicted large value given by the quantum field theory.
\par 
The second approach aims at modifying the geometry of spacetime, i.e. the Einstein's gravity in the general theory of relativity (GR) at large distances, specifically beyond our Solar System to produce accelerating cosmological solutions~\cite{Carroll:2003wy, Cognola:2007zu,Clifton:2011jh}. This has given rise to the concept of "modified gravity", with numerous theories available in the literature. Extensive reviews in modified gravity theories can be found in the Refs.~\cite{Nojiri:2017ncd, Nojiri:2006ri, Capozziello:2019cav}. Many of the theories of modified gravity aims at modifying the linear function of scalar curvature, $R$ responsible for the Einstein tensor in the Einstein equations of GR. So it is obvious that the alterations are brought about in such a way so as to generalize the gravitational Lagrangian which takes a special form $\mathcal{L}_{\rm GR}= R$ in case of GR. An extensively studied theory in this context is the $f(R)$ gravity where the gravitational Lagrangian $\mathcal{L}_{\rm GR}=R$ is replaced by an analytic function of $R$, i.e., $\mathcal{L}_{f(R)}=f(R)$. Via this generalization, we can explore the non-linear effects of the scalar curvature $R$ in the evolution of the universe by choosing a suitable function for $f(R)$. In this specific case, extensive reviews on this theory are available in the Refs.~\cite{DeFelice:2010aj, Sotiriou:2008rp}.
\par 
From a theoretical point of view, the viability of $f(R)$ dark energy models have been studied in \cite{PhysRevLett.98.131302}. In this paper the authors ruled out the $f(R)$ theories where a power of $R$ is dominant at
large or small $R$. The effects of a non-minimal curvature-matter coupling was studied in \cite{Azizi:2014qsa}, and constraints of the $f(R)$ dark energy models were derived in \cite{Amendola:2006we}. In Ref.~\cite{Sotiriou:2006hs}, the author studied the interplay between scalar–tensor theory and $f(R)$ theories of gravity considering the Palatini formalism. A specific scheme for $f(R)$ reconstruction was developed recently \cite{Nojiri:2006gh}, while large scale structure of $f(R)$ gravity was investigated in~Ref.~\cite{Song:2006ej}. Moreover, various papers have investigated different aspects of the latter theory by considering various techniques \cite{Shah:2019mxn,Odintsov:2017tbc, Bamba:2014wda}, and a survey of the generic f(R) models in various formulations is carried out in \cite{Capozziello:2009nq}. Further generalizations to the $f(R)$ modified theories of gravity have been proposed by introducing some couplings between the geometrical quantities and the matter sector. One interesting model is the one where the Lagrangian is constructed by considering a generic function of the Ricci scalar $R$ and of the trace of the stress-energy tensor $T$. Such modifications gave rise to $f(R,T)$ theories \cite{Harko:2011kv,Arik:2013sti, Nagpal:2019vre}. Moreover since scalar fields play a fundamental role in cosmology, $f(R,T^{\phi})$ theories were proposed by Harko et al in Ref.~\cite{Harko:2011kv}, where $T^{\phi}$ is the trace of the stress energy of the scalar field. A different type of coupling between geometry and matter was proposed \cite{Haghani:2013oma}, the generic $f(R,T,R_{\mu\nu}T^{\mu\nu})$ gravity theory. This is a more generic gravity theory in which matter is nonminimally coupled to geometry, the Lagrangian corresponding to the gravitational field has a general dependence of the Ricci scalar, the trace of the matter energy-momentum tensor, and the contraction between the Ricci tensor and the matter energy-momentum tensor. Further, in~\cite{Bertolami:2007gv,Harko:2010zi}, a model with a non-minimally coupling between the density Lagrangian matter and the curvature $R$ was introduced. In this model, the Lagrangian corresponds to $f_1(R)+(1+\lambda f_2(R))L_{\rm m}$ where $\lambda$ is a constant and $L_{\rm m}$ is the matter Lagrangian density. Within these models in which the matter field is non-minimally coupled to gravity any particle is subject to an extra force appearing in a direction which is orthogonal to the four-velocity~\cite{Bertolami:2007gv}. The latter model proposed in \cite{Bertolami:2007gv} was extended to the case of generic couplings to both matter and gravity in a recent paper \cite{Harko:2008qz}, considering a specific Lagrangian given by  $f_1(R)+G(L_{\rm m})f_2(R)$. Furthermore, for the non-minimal geometry coupling models the Palatini formulation has been proposed in \cite{Harko:2010hw}. In this context, a further specific extension related to the latter gravity theories was proposed in \cite{Harko:2010mv} by embedding into the Lagrangian an arbitrary function of the $f(R,L_{\rm m})$ type.
\par 
More comprehensive ideas and reviews on modified gravity theories from different points of view has been considered in \cite{Nojiri:2017ncd, DeFelice:2010aj, DAgostino:2018ngy}. The advances in the recent  cosmology using a dynamical system approach in dark energy and modified gravity theories have appeared recently \cite{Bahamonde:2017ize}. In continuation of the above generalization procedure for the $f(R,L_{m})$ theory we can also choose to modify the corresponding Lagrangian by including some analytic function of $T_{\mu\nu}T^{\mu\nu}$, where $T_{\mu\nu}$ is the stress energy-momentum tensor of the matter component. Hence, such a choice of the corresponding Lagrangian will give rise to $f(R,T_{\mu\nu}T^{\mu\nu})$ theories of gravity. It should be kept in mind that in such a scenario, we are not introducing new forms of non-linear fluid stresses \cite{Barrow:1990vx, Nojiri:2005sr} like the scalar field, bulk viscosity \cite{Barrow:1988yc}, or Chaplygin gas \cite{Bento:2002ps, Gorini:2002kf}. In 2014, Katirci and Kavuk ~\cite{Arik:2013sti} proposed such a theory for the first time, a covariant generalization of GR which allows the existence of a term proportional to $T_{\mu\nu}T^{\mu\nu}$ in the action functional. Further studies on this theory was carried out ~\cite{Board:2017ign, Roshan:2016mbt}, where specific models of this gravity theory have been considered. Roshan et al ~\cite{Roshan:2016mbt} analyzed the possibility of a bounce at early times within the energy momentum squared gravity (EMSG) model, with the specific functional given by $f(R,T^2)=R+\eta T^2$, where $\eta$ is a constant. Board and Barrow \cite{Board:2017ign} studied the cosmology of the energy momentum powered gravity (EMPG) model which is a generalization of the EMSG theory, where the model is characterized by $f(R,T^2)=R+\eta (T^2)^n$, with $\eta$ and $n$ constant parameters. Non exotic matter wormholes are studied in the framework of EMSG in~\cite{Moraes:2017dbs}, and possible constraints from neutron stars were discussed in~\cite{Akarsu:2018zxl}. Furthermore, recent studies \cite{Nari:2018aqs,Akarsu:2018aro,Keskin:2018bkg} have considered various cosmological applications of the energy momentum squared gravity theory. It has been shown ~\cite{Liu:2016qfx} that the quantum fluctuations associated to the metric tensor can produce additional cross terms between the Ricci and the energy-momentum tensor. In the framework of energy momentum squared gravity theories, the late time acceleration of the Universe have been investigated in~\cite{Akarsu:2017ohj}, considering the case of a pressure-less fluid. In this specific case the authors~\cite{Akarsu:2017ohj} constrained different parameters of the corresponding model by relying on various values of the Hubble parameter. 
\par 
From the above handful of literature it is clear that $f(R,T_{\mu\nu}T^{\mu\nu})$ gravity along with its EMSG and EMPG specializations need more attention and hence motivations to study such theories are quite high. As both the concepts of dark energy and modified gravity are till date inadequate to properly explain the observations, it is obvious that there is a lot of room for improvement in both the sectors and there is an open invitation to work and improve both the territories. This work is one such novel attempt to improve upon our existing knowledge of modified gravity theories. The present work will focus on some generalizations of the EMSG model of modified gravity and will try to explore its features via a dynamical system analysis.
\par 
The paper is organized as follows: Section II deals with cosmology of energy momentum squared gravity theories. In section III we give a detailed analysis of the dynamical system for two specific case of gravity functions. Finally the paper ends with the corresponding concluding remarks in section IV.

\section{Energy-momentum Squared Cosmology}

The action of our model can be written as~\cite{Board:2017ign}
\begin{equation}
S=\frac{1}{2\kappa^2}\int d^4x \sqrt{-g} f(R,\mathbf{T^2})  + S_{\rm m},
\end{equation} 
where $f$ is a function depending on the square of the energy-momentum tensor $\mathbf{T^2}=T^{\mu\nu}T_{\mu\nu}$ and the scalar curvature $R$. Here, $\kappa^2=8\pi G$ and $S_{\rm m}$ represents the action corresponding to the matter component.

If we vary the action with respect to the metric we arrive at the following field equations
\begin{equation}
R_{\mu \nu}f_R  +g_{\mu\nu} \Box f_R-\nabla_{\mu}\nabla_{\nu} f_R-\frac{1}{2} g_{\mu\nu}f=\kappa^2 T_{\mu \nu}-f_{\mathbf{T^2}} \Theta_{\mu \nu}\,, \label{FieldEq}
\end{equation}
where $\Box=\nabla_\mu \nabla^\mu$, $f_{R}=\partial f/\partial R$, $f_{\mathbf{T^2}}=\partial f/\partial \mathbf{T^2}$ and
\begin{equation}
    \Theta_{\mu\nu}=\frac{\delta (\mathbf{T^2})}{\delta g^{\mu\nu}}= \frac{\delta (T^{\alpha\beta}T_{\alpha\beta})}{\delta g^{\mu\nu}}=-2L_{\rm m}\Big(T_{\mu\nu}-\frac{1}{2}g_{\mu\nu}T\Big)-T\, T_{\mu\nu}+2T^{\alpha}_{\mu}T_{\nu\alpha}-4T^{\alpha\beta}\frac{\partial^2 L_{\rm m}}{\partial g^{\mu\nu}\partial g^{\alpha\beta}}\,,\label{Theta}
\end{equation}
where $T$ is the trace of the energy-momentum tensor. By taking covariant derivatives with respect to the field equation~\eqref{FieldEq}, one finds the following conservation equation
\begin{eqnarray}
\kappa^2\nabla^\mu T_{\mu\nu}=-\frac{1}{2}g_{\mu\nu}\nabla^\mu f+\nabla^\mu(f_{\mathbf{T}^2}\Theta_{\mu\nu})\,.\label{conservation1}
\end{eqnarray}
As one can see from the above equation that in general, the conservation equation does not hold for this theory. If one chooses $f(R,\mathbf{T}^2)=2\alpha \log(\mathbf{T}^2)$, one gets the same result reported in \cite{Akarsu:2019ygx}.

In the following, we will concentrate on the flat FLRW cosmology for this model whose metric is described by
\begin{equation}
ds^2=-dt^2+a^2(t)\delta_{ik}dx^idx^k,
\end{equation}
with $\delta_{ik}$ being the Kronecker symbol and $a(t)$ the scale factor. Let us now consider that the matter content is described by a standard perfect fluid with $T_{\mu\nu}=(\rho+p)u_\mu u_\nu + p g_{\mu\nu}$ with $u_{\mu}$ being the 4-velocity and $\rho$ and $p$ are the energy density and the pressure of the fluid respectively. Using these, the energy-momentum tensor gives us $T^2=\rho^2+3p^2$. Further, let us assume $L_{\rm m}=p$ which allows us to rewrite $\Theta_{\mu\nu}$ defined in eqn. \eqref{Theta} as a quantity which does not depend on the function $f$, namely~\cite{Board:2017ign}
\begin{equation}
    \Theta_{\mu\nu}=-\Big(\rho^2+4 p\rho+3p^2\Big)u_\mu u_\nu\,.
\end{equation}
The modified FLRW equations which corresponds to this particular action are given by
\begin{eqnarray}
-3f_R\Big(\dot{H}+ H^2\Big)+\frac{f}{2}+3 H \dot{f_R}&=&\kappa^2\Big(\rho+\frac{1}{\kappa^2}f_{\mathbf{T^2}}\Theta^2\Big)\,,\label{FW1} \\
-f_R(\dot{H}+3 H^2)+\frac{1}{2}f+\ddot{f_R}+2 H \dot{f_R}&=&-\kappa^2 p\,,\label{FW2}
\end{eqnarray}
where dots denote differentiation with respect to the cosmic time $t$, $H=\dot{a}/a$ is the Hubble parameter, and 
\begin{equation}
\mathbf{\Theta^2}:=\Theta_{\mu\nu}\Theta^{\mu\nu}=\rho^2+4p \rho+3p^2\label{Theta2}
\end{equation}
was defined. The conservation equation \eqref{conservation1} reads as follows
\begin{eqnarray}
\kappa^2(\dot{\rho}+3H(\rho+p))&=&-\mathbf{\Theta^2}\dot{f}_{\mathbf{T^2}}-f_{\mathbf{T^2}} \Big[3 H\mathbf{\Theta^2}+\frac{d}{dt}\Big(2\rho p+\frac{1}{2}\mathbf{\Theta^2}\Big)\Big]\,.\label{conservation3}
\end{eqnarray}
Clearly, the standard conservation equation does not hold in $f(R,\mathbf{T^2})$ cosmology for an arbitrary function. If one chooses $f(R,\mathbf{T^2})=f(R)$, all the terms on the RHS of the above equation are zero and the standard conservation equation is recovered. In the following, a standard barotropic equation of state will be assumed:
\begin{equation}
    p=w \rho\,,
\end{equation}
where $w$ is the equation of state parameter. Using this relation one gets that
\begin{equation}
    \mathbf{\Theta^2}=(1+4w+3w^2)\rho^2\,,
\end{equation}
and then the conservation equation \eqref{conservation3} becomes
\begin{eqnarray}
\dot{\rho}+3H(w+1)\rho&=&-f_{\mathbf{T^2}} \left[3 \left(3 w^2+4 w+1\right) H \rho ^2+\left(3 w^2+8 w+1\right) \rho \dot{ \rho}\right]\nonumber\\
&&-\left(3 w^2+4 w+1\right) \rho^2 \dot{f}_{\mathbf{T^2}}\,.\label{conservation4}
\end{eqnarray}
In the next section, the dynamical system of the general model will be found and then some specific forms of the function will be adopted to analyse its cosmological properties by using dynamical system techniques. Hereafter, we will use geometric units such that $\kappa^2=1$.

\section{Dynamical system}
In this section we will derive the general form of the dynamical system for the modified FLRW equations described by Eqs.~\eqref{FW1}-\eqref{FW2}. Let us first introduce the following dimensionless variables
\begin{equation}
x_1=\frac{\dot{f}_{R}}{f_{R} H}\,,\quad x_2=\frac{f}{6 H^2f_{R}}\,,\quad x_3=\frac{R}{6 H^2}\,,\quad x_4=\frac{\rho }{3 H^2 f_R}\,,\quad x_5=\left(3 w^2+4 w+1\right) \rho f_{\mathbf{T^2}}\,.
\end{equation}
Using these quantities one finds that the first Friedmann equation given by~\eqref{FW1} becomes
\begin{eqnarray}
x_3+x_4+x_4 x_5-x_1-x_2=1\,.\label{constraint}
\end{eqnarray}
The dynamical system for these five dimensionless variables become
\begin{eqnarray}
\frac{dx_1}{dN}&=&\Gamma -x_1\left(x_1+\Psi \right)\,,\label{Eq1}\\
\frac{dx_2}{dN}&=&\Xi -2 x_2 \Psi-x_1x_2\,,\\
\frac{dx_3}{dN}&=&\mho -2 x_3\Psi\,,\\
\frac{dx_4}{dN}&=&\Pi -2 x_4 \Psi -x_1 x_4\,,\\
\frac{dx_5}{dN}&=&3 \left(3 w^2+4 w+1\right) \Delta \,   x_4+\frac{\Pi x_5}{x_4}\,,\label{Eq5}
\end{eqnarray}
where $N=\log(a)$ and we have further defined the following parameters
\begin{equation}
\mho=\frac{\dot{R}}{6 H^3}\,,\quad \Xi =\frac{\dot{f}}{6 f_R H^3}\,,\quad \Psi= \frac{\dot{H}}{H^2}\,,\quad \Delta =f_R H \dot{f}_{\mathbf{T^2}}\,,\quad \Pi =\frac{\dot{\rho}}{3 f_R H^3}\,.\label{auxiliary}
\end{equation}
On the other hand, the conservation equation \eqref{conservation4} can be written as follows ($w\neq-1$ and $w\neq-1/3$)
\begin{eqnarray}
\Delta = -\frac{\Pi \Big((w (3 w+8)+1) x_5+(w+1) (3 w+1)\Big)+3 (w+1) (3 w+1) x_4 \left(x_5+w+1\right)}{3 (w+1)^2 (3 w+1)^2 x_4{}^2}\,.\label{Deltais}
\end{eqnarray}
Then, one can further notice the following extra relationships
\begin{eqnarray}
\Psi= x_3-2\,,\quad
\Xi =\frac{\left(3 w^2+1\right) \Pi  x_5}{3 w^2+4 w+1}+\mho\,,
\end{eqnarray}
and then rewrite the dynamical system~\eqref{Eq1}-\eqref{Eq5} as follows
\begin{eqnarray}
\frac{dx_2}{dN}&=&\frac{\left(3 w^2+1\right) \Pi  x_5}{3 w^2+4 w+1}+x_2^2-\left(3 x_3+x_4 \left(x_5+1\right)-5\right) x_2+\mho \,,\label{dx2}\\
\frac{dx_3}{dN}&=&\mho -2 \left(x_3-2\right) x_3\,,\label{dx3}\\
\frac{dx_4}{dN}&=&\Pi -x_4 \left(-x_2+3 x_3+x_4+x_4 x_5-5\right)\,,\label{dx4}\\
\frac{dx_5}{dN}&=&-\frac{\Pi  \left(4 w x_5+3 w^2+4 w+1\right)+3 \left(3 w^2+4 w+1\right) x_4 \left(x_5+w+1\right)}{(w+1) (3 w+1) x_4}\,,\label{dx5}
\end{eqnarray}
where we have also used the Friedmann constraint~\eqref{constraint} and the conservation equation~\eqref{Deltais}. Furthermore, in terms of dimensionless and auxiliary intermediate variables, the Friedmann acceleration equation~\eqref{FW2} reduces to
\begin{equation}
   x_3=\Gamma +3 w x_4+2 x_1+3 x_2-1\,.
\end{equation}
To close the dynamical system, one needs to impose a specific form of the function $f$. By doing this, one needs to derive how the parameters $\Pi$ and $\mho$ are either constants or depend on the dimensionless variables $x_2,x_3,x_4$ and $x_5$. Lastly, it can be seen that this particular choice of dimensionless variables does not necessarily imply that the effective matter density parameter needs to satisfy the usual standard existence conditions, i.e. $0 \leq \Omega_{\rm m}^{\rm eff}=x_4 \leq 1$ due to the appearance of the variation of the gravity functional with respect to the scalar curvature. However, in our analysis we shall consider the necessity of the  standard existence conditions as a basic requirement for the validity of the corresponding critical points due to the  complexity of the phase space.

Let us finish this section by noting that the effective state parameter is only related to the dimensionless parameter $x_3$ as follows
\begin{equation}
    w_{\rm eff}=-1-\frac{2}{3}  \left(x_3-2\right)\,.\label{weff}
\end{equation}
This expression will be used to understand the nature of the critical points.
\subsection{Case 1: $f(R,\mathbf{T^2})=f_0R^n(\mathbf{T^2})^m$}If we specify the following functional $f(R,\mathbf{T^2})=f_0 R^n (\mathbf{T^2})^m$, where $m,n$ and $f_0$ are constant parameters, then we can close the dynamical system. By using the definition of $\Delta$ for this specific functional, we can deduce the following intermediate equation:
\begin{equation}
    \Delta = \frac{2 m n x_2^2 }{3 \left(3 w^2+1\right) x_3^2 x_4^3}\Big(2 (m-1) \Pi  x_3+n x_4 \mho \Big)\,,\label{Deltaiss}
\end{equation}
which represents an interrelation between the definition of $\Delta$ in the conservation equation and the intermediate variables $\mho$ and $\Pi$.
Moreover, in this specific case, one can note that we have an interrelation between the second and third dimensionless variables, namely, \begin{equation}
\label{hhhhh}
    x_3= n x_2\,.
\end{equation}
Furthermore, for this specific model, one can write the following:
\begin{equation}
   x_2(n-1)+x_4+x_4 x_5-1= \frac{(n-1) \mho}{nx_2}+\frac{2 m \Pi  }{ x_4}\,,\label{ola}
\end{equation}
where we have used the constraint~\eqref{constraint}. Thus, we have three equations~\eqref{Deltais},~\eqref{Deltaiss} and \eqref{ola} for the variables $\mho,\Delta$ and $\Pi$ that can be solved in terms of the dimensionless variables $x_2,x_4$ and $x_5$, yielding
\begin{eqnarray}
 \Delta&=&\frac{1}{3x_4\tilde{G}}\Big[ 2 m n x_2^2 \Big(6 m n^2 (w+1) (3 w+1) x_2 \left(w+x_5+1\right)+x_3 \left(n x_1 \left(3 w^2+(w (3 w+8)+1) x_5+4 w+1\right)\right.\nonumber\\
&&\left.-6 (m-1) (n-1) (w+1) (3 w+1) \left(w+x_5+1\right)\right)\Big) \Big]\,,\\
\Pi&=&-\frac{1}{\tilde{G}}\Big[ (w+1) (3 w+1) x_3 x_4 \left(2 m n^2 (w+1) (3 w+1) x_1 x_2^2+3 (n-1) \left(3 w^2+1\right) x_3 x_4 \left(w+x_5+1\right)\right) \Big]\,,\\     
\Xi&=&\frac{1}{\tilde{G}}\Big[ x_3 \left(3 \left(3 w^2+1\right) x_3 x_4 \left(w+x_5+1\right) \left(2 m n (w+1) (3 w+1) x_2-(n-1) \left(3 w^2+1\right) x_4 x_5\right)\right.\nonumber
    \\&&+x_1 \left(2 m n (w+1) (3 w+1) x_2^2 \left(2 (m-1) (w+1) (3 w+1) x_3-n \left(3 w^2+1\right) x_4 x_5\right)\right.
    \nonumber\\&&\left.+\left(3 w^2+1\right) x_3^2 x_4 \left(3 w^2+(w (3 w+8)+1) x_5+4 w+1\right)\left.\right)\right)\Big]\,,\\ 
\mho&=&\frac{1}{\tilde{G}}\Big[ x_3^2 \Big(6 m n (w+1) (3 w+1) (3 w^2+1) x_2 x_4 (w+x_5+1)+x_1 (4 (m-1) m n (w+1)^2 (3 w+1)^2 x_2^2
    \nonumber\\&&+(3 w^2+1) x_3 x_4 (3 w^2+(w (3 w+8)+1) x_5+4 w+1)\Big) \Big]\,,
\end{eqnarray}
where for simplicity, we have defined the quantity
\begin{eqnarray}
    \tilde{G}&\equiv&(n-1) x_3 \left(4 (m-1) m n (w+1)^2 (3 w+1)^2 x_2^2+\left(3 w^2+1\right) x_3x_4 \left((w (3 w+8)+1) x_5+3 w^2+4 w+1\right)\right)\nonumber\\
    &&-4 m^2 n^3 (w+1)^2 (3 w+1)^2 x_2^3\,.
\end{eqnarray}
Therefore, the dynamical system~\eqref{dx2}-\eqref{dx5} is reduced to a 3 dimensional one, given by
\begin{eqnarray} 
\frac{dx_2}{dN}&=&x_2 \Big((1-3 n) x_2-x_4 \left(x_5+1\right)+5\Big)-\frac{1}{G}\Big[3 \left(3 w^2+1\right) x_4\left(x_5+w+1\right) \Big((n-1) \left(3 w^2+1\right) x_4 x_5\nonumber\\
&&-2 m n (w+1) (3 w+1) x_2\Big)-x_2 \Big((n-1) x_2+x_4 \left(x_5+1\right)-1\Big) \Big(2 m (w+1) (3 w+1) \Big\{-n \left(3 w^2+1\right) x_4 x_5\nonumber\\
&&+2 (m-1) n (w+1) (3 w+1) x_2\Big\}+n \left(3 w^2+1\right) x_4 \Big\{(w (3 w+8)+1) x_5+3 w^2+4 w+1\Big\}\Big)\Big]\,,\\
    \frac{dx_4}{dN}&=&x_4 \Big[-3 n x_2+x_2-x_4-x_4 x_5+5-\frac{(w+1) (3 w+1)}{G}\Big\{2 m n (w+1) (3 w+1) x_2 \Big((n-1) x_2+x_4 \left(x_5+1\right)-1\Big)\nonumber\\
    &&+3 (n-1) \left(3 w^2+1\right) x_4 \left(x_5+w+1\right)\Big\}\Big]\,,\\
\frac{dx_5}{dN}&=&\frac{1 + 4 w + 3 w^2}{G}\Big[2 m \left(3 w^2+4 w+1\right) x_2 \Big(6 m w+6 m+(n-1) n x_2+n x_4+6 n w+5 n-6 w-6\Big)\nonumber\\
&&+x_4 x_5^2 \Big(8 m n w x_2-3 (n-1) \left(3 w^2+1\right)\Big)+8 m (n-1) n w x_2^2-3  x_4(n-1) \left(3 w^3+3 w^2+w+1\right)\nonumber\\
&&+2x_2x_4 m n \left(3 w^2+8 w+1\right)+4 m x_2 \Big(3 m \left(3 w^2+4 w+1\right)+n \left(9 w^2+10 w+3\right)-3 \left(3 w^2+4 w+1\right)\Big)\Big]\,,
\end{eqnarray}
 where again, for simplicity, we have defined the function
 \begin{eqnarray}
     G&\equiv&(n-1) \Big[4 (m-1) m (w+1)^2 (3 w+1)^2 x_2+\left(3 w^2+1\right) x_4 \left((w (3 w+8)+1) x_5+3 w^2+4 w+1\right)\Big]\nonumber\\
     &&-4 m^2 n (w+1)^2 (3 w+1)^2 x_2\,.
 \end{eqnarray}

\begin{table}[H]
\centering
 \begin{tabular}{||c | c | c | c||} 
 \hline
 Cr.P. & $x_2$ & $x_4$ & $x_5$ \\ [0.5ex] 
 \hline\hline
 $P_1$ & $\frac{m (n+5)+4 n-5}{(m+2) n^2+(2 m-3) n-m+1}$ & 0 & $\frac{-3 m n (n+2)+3 m+n (11-7 n)-3}{3 m (n (n+2)-1)+6 n^2-9 n+3}$ \\ [0.5ex]
 \hline
 $P_2$ & $\frac{52 m^2 n-12 m^2+56 m n^2-52 m n+12 m+16 n^3-28 n^2+13 n-3+(1-2 m)\tau}{4 n^2 (2 m+n-1) (2 m+2 n-1)}$  & $\frac{-3+m (6-26 n)+(13-8 n) n+\tau}{4 n^2}$ & $-\frac{3+6 m (4 m-3)-13 n+26 m n+8 n^2+\tau}{6 (2 m+n-1) (2 m+2 n-1)}$ \\[0.5ex]
 \hline
 $P_3$ & $\frac{52 m^2 n-12 m^2+56 m n^2-52 m n+12 m+16 n^3-28 n^2+13 n-3-(1-2 m)\tau}{4 n^2 (2 m+n-1) (2 m+2 n-1)}$ & $\frac{-3+m (6-26 n)+(13-8 n) n-\tau}{4 n^2}$ & $\frac{-26 m n+6 m (3-4 m)-8 n^2+13 n-3+\tau}{6 (2 m+n-1) (2 m+2 n-1)}$ \\[0.5ex]
 \hline
  \end{tabular}
 \caption{The structure of the phase space for the $f(R,\mathbf{T^2})=f_0R^n(\mathbf{T^2})^m$ model. We have defined the following quantity:  $\tau=\sqrt{4 m^2 (n (49 n-78)+9)+4 m (n (7 n (8 n-19)+78)-9)+(n (8 n-13)+3)^2}$.} \label{TableMihai}
 \end{table}
 \par 
 In the following, we will only focus on the dust case $(w=0)$. In this case, the structure of the phase space consists of three critical points which are valid from a physical point of view  described in Table~\ref{TableMihai}. Our analysis considered the requirement of the standard existence conditions which implies that the critical points are located in the real space, while the effective matter density parameter is non--negative and satisfies $0\leq \Omega_{\rm m}^{\rm eff} \leq 1$. The latter requirement represents an extra condition \cite{Carloni:2015lsa} added to the existence conditions due to the complexity of the phase space, by generalizing the usual conditions considered in the minimal coupling case. As can be observed from the Table~\ref{TableMihai}, the structure of the phase space is sensitive to the values of the $m$ and $n$ parameters. Furthermore, the stability criteria for each critical point is analyzed in detail by determining the corresponding eigenvalues for each type of solution, constraining the possible values of the $m$ and $n$ parameters from a dynamical point of view. Due to the high complexity of the corresponding eigenvalues for the $P_2$ and $P_3$ critical points, the analysis relies on numerical evaluations which are described in figures. Each critical points have the following features: 
 \begin{itemize}
    \item \textit{Point $P_1$}:
 The first critical point $P_1$ represents a solution dominated by the geometrical dark energy component, with physical effects from the shape of the energy momentum squared gravity functional $f(R,\mathbf{T^2})$ and its variation with respect to the matter energy momentum tensor $\mathbf{T^2}$. In this case the matter density parameter $\Omega_{\rm m}^{\rm eff}$ is equal to zero, corresponding to an cosmological era dominated completely by the geometrical dark energy component. This critical point always exist, the only constraint come from the requirement that the solution has a non--zero denominator, e.g. resulting in: $(m+2) n^2+(2 m-3) n-m+1\neq0$ and also $3 m (n (n+2)-1)+6 n^2-9 n+3\neq0$. Using the definition of the effective equation of state in Eq. \eqref{weff} one can find that at this critical point, the dynamical features depends heavily on the values the $m,n$ parameters, considering also the Eq. \eqref{hhhhh}: 
 \begin{equation}
     w_{\rm eff}=\frac{-m n^2-8 m n-m-6 n^2+7 n+1}{3 \left(m n^2+2 m n-m+2 n^2-3 n+1\right)}\,.
 \end{equation}
 The stability of this solution is also affected by the $m$ and $n$ parameters, as can be seen from its corresponding eigenvalues:
 \begin{equation}
    \Big[-3,-\frac{m n+5 m+4 n-5}{m+n-1},\frac{2 m n^2-4 m n-n^2+2 n}{m n^2+2 m n-m+2 n^2-3 n+1} \Big]\,.
 \end{equation}
\par 
The characterization of this critical point is displayed in the Fig. \ref{fig:awage1} where we have shown the physical effects of the $m$ and $n$ parameters, obtaining some possible constraints which corresponds to the stable case. The effective equation of state is sensitive to the variation of the $m$ and $n$ constants and the solution can manifest a large spectrum of dynamical eras, starting from a super--accelerated expansion to matter, stiff and super--stiff cosmological epochs. Due to the large spectrum of physical epochs present, we have displayed in the right panel of Fig. \ref{fig:awage1} the variation of the effective equation of state only in regions which corresponds to intervals of interest for modern cosmology. However, for some values of the $m$ and $n$ parameters, this point can represent either a super-accelerating late-time attractor or a saddle matter dominated point.

\begin{figure}[!ht]
\minipage{0.5\textwidth}
\subfigure{
  \includegraphics[height=0.8\linewidth]{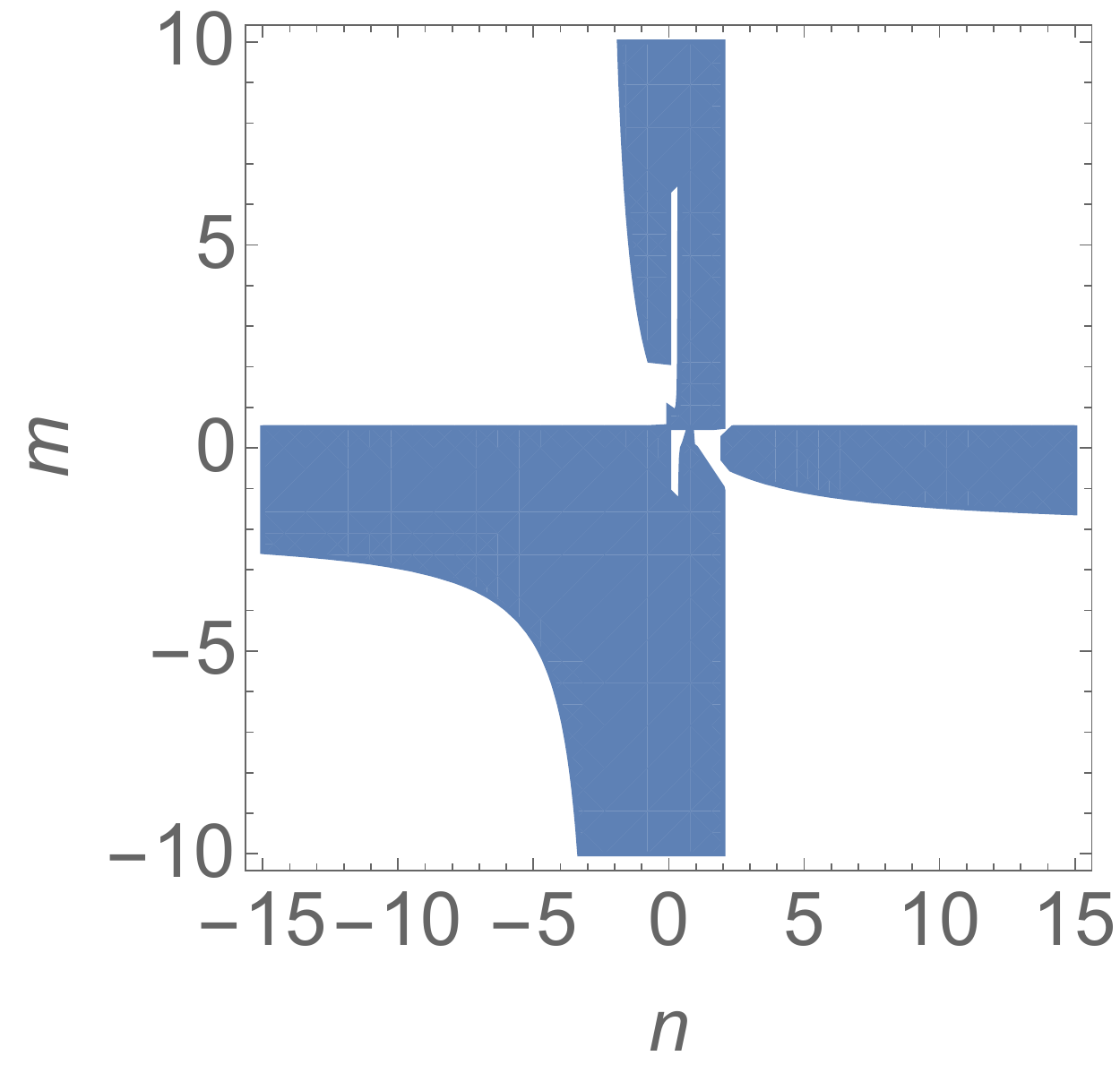}
  {a)}\label{fig:kl0}}
\endminipage\hfill
\minipage{0.5\textwidth}
  \subfigure{\includegraphics[height=0.8\linewidth]{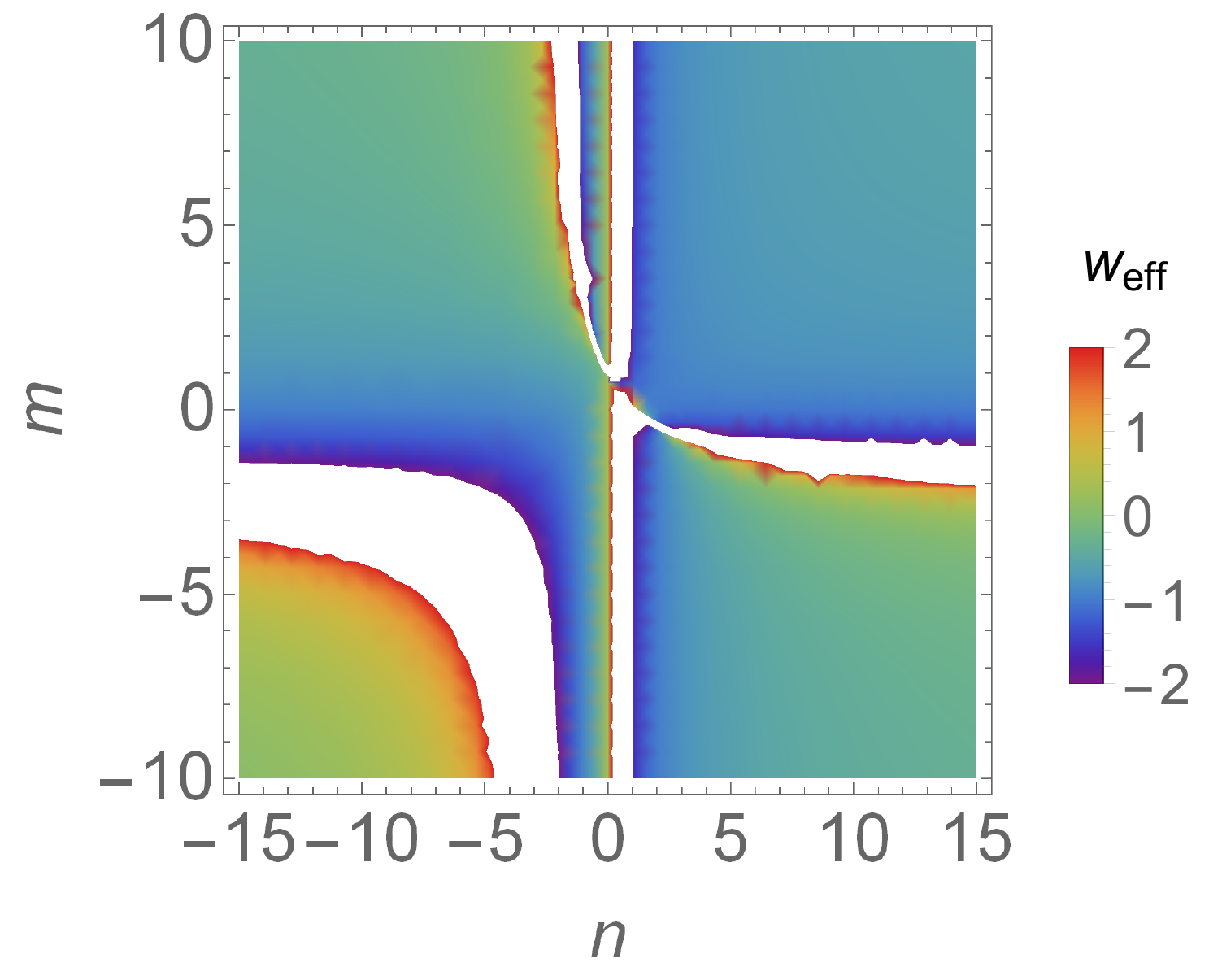}
  {b)}\label{fig:kl1}}
\endminipage
\caption{a) The regions where the critical point $P_1$ is stable;  (the white regions corresponds to saddle dynamical behavior) b) The variation of the effective equation of state for the $P_1$ solution; (the white regions corresponds to an interval larger than $[-2,+2]$).}\label{fig:awage1}
\end{figure}

\par 
\item \textit{Point $P_2$}: The second critical point $P_2$ represents a solution characterized by the physical effects coming from the stress energy momentum coupling, the shape of the functional $f$, and matter energy density. In this case, the effective matter density is equal to 
\begin{equation}
    \Omega_{\rm m}^{\rm eff}=\frac{-3+m (6-26 n)+(13-8 n) n+\tau}{4 n^2}\,,
\end{equation}
showing a sensitivity to the values of the $m$ and $n$ constants. Analyzing the stability criteria in this case, the expression of the corresponding eigenvalues are too cumbersome to be written in the manuscript. Hence, we shall analyze the physical implications of the $P_2$ solution by relying only on numerical evaluations. The numerical description of the second critical point is shown in Fig.~\ref{fig:awage2} where it can be seen that this critical point represents an epoch characterized by an accelerated or super--accelerated expansion. Note that in this case we have taken into account the standard existence condition for the effective matter density parameter which implies $0\leq \Omega_{\rm m}^{\rm eff} \leq 1$. Hence, depending on the values of the $m$ and $n$ parameters, we can obtain a stable solution characterized by an accelerated expansion, in agreement with  the current evolution of the known Universe.

\begin{figure}[!ht]
\minipage{0.5\textwidth}
  \subfigure{\includegraphics[height=0.8\linewidth]{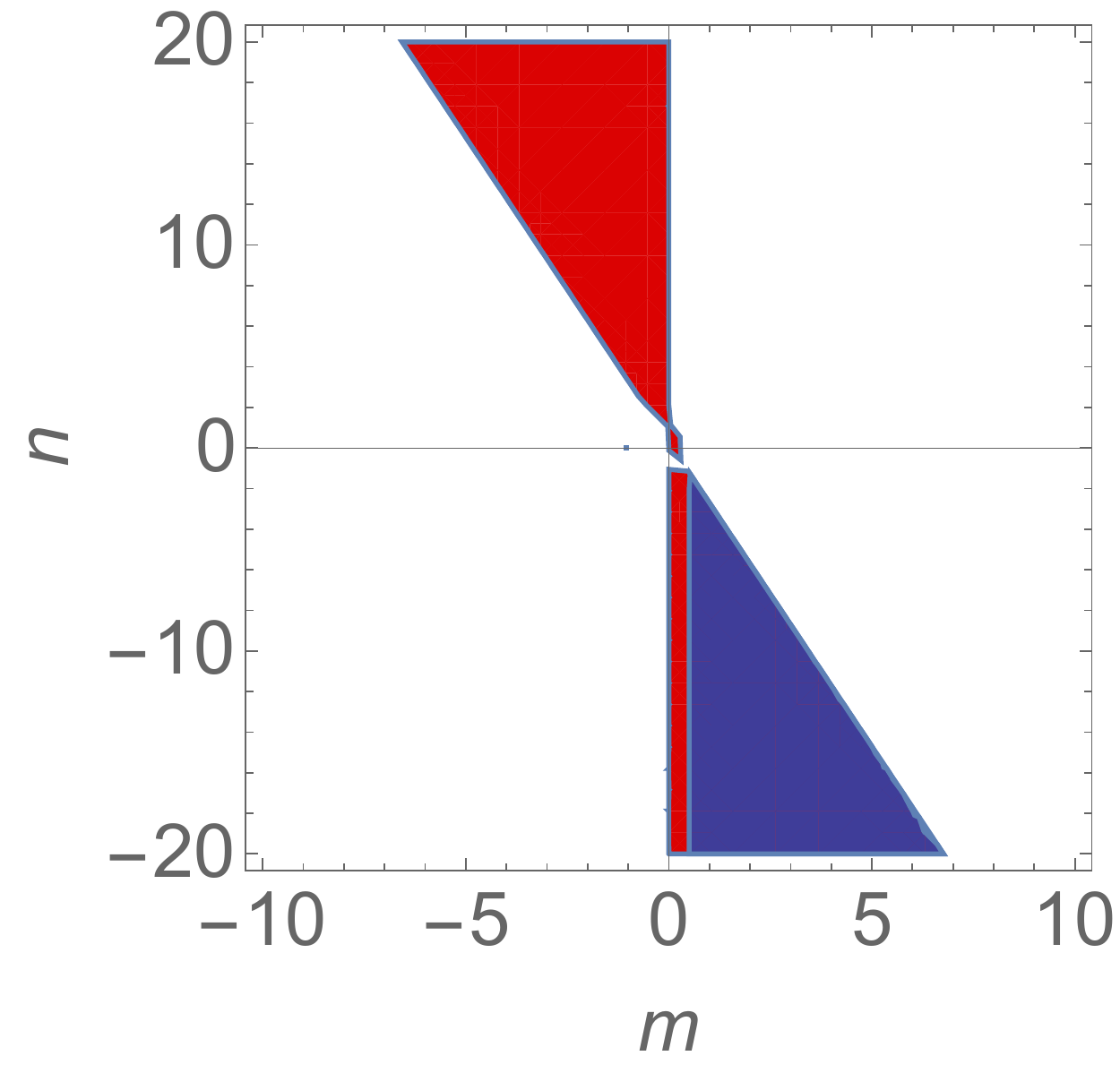}
  {a)}\label{fig:kl0}}
\endminipage\hfill
\minipage{0.5\textwidth}
  \subfigure{\includegraphics[height=0.8\linewidth]{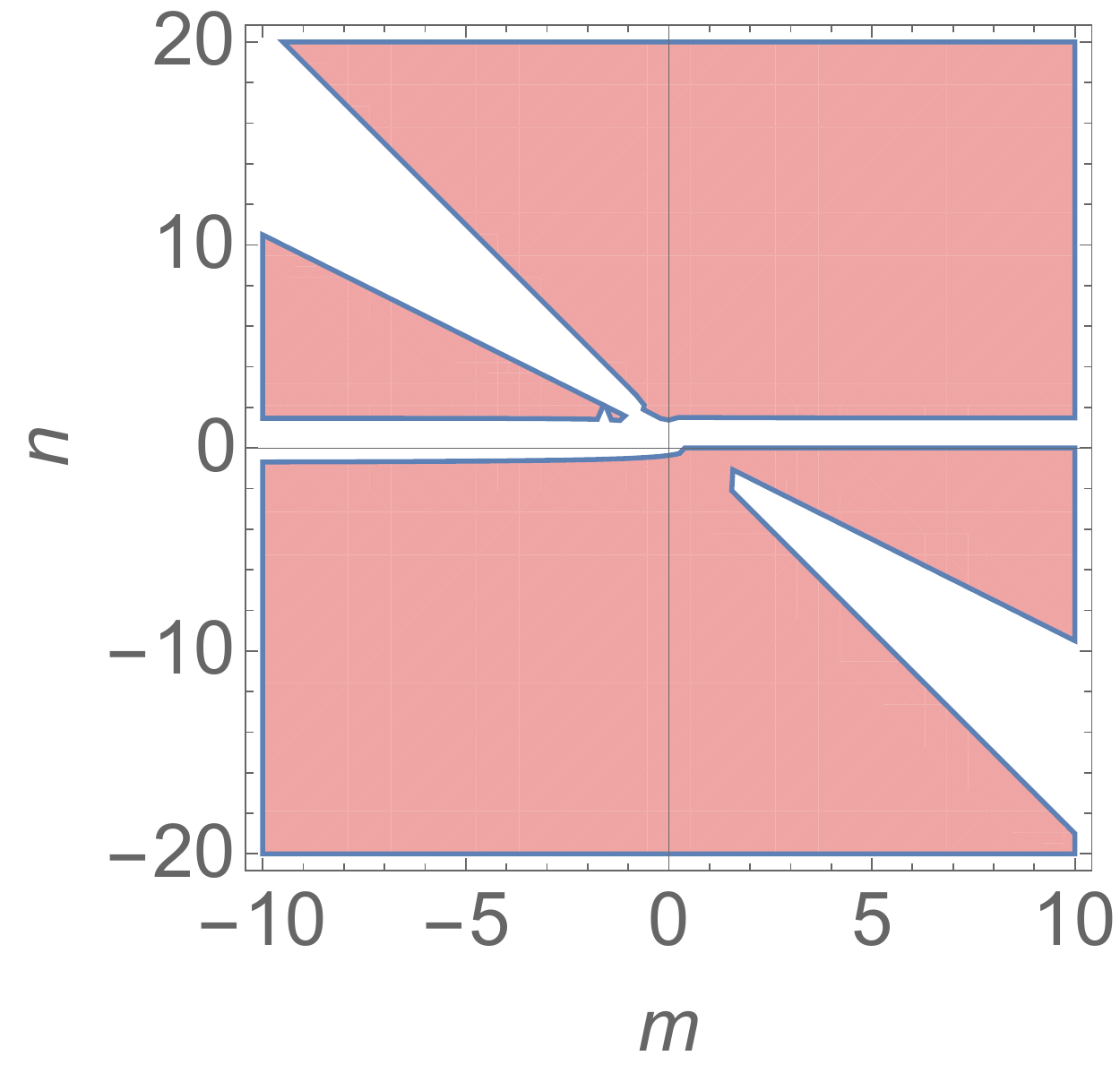}
  {b)}\label{fig:kl1}}
\endminipage\hfill\\
\caption{a) The figure shows the regions where the standard existence condition associated to the $P_2$ critical point is satisfied (red); the  stable intervals which includes the existence regions (magenta)  b) The limited regions corresponding to the acceleration intervals where $w_{\textbf{eff}}< -1/3$. }\label{fig:awage2}
\end{figure}

\begin{figure}[!h]
\includegraphics[height=0.4\linewidth]{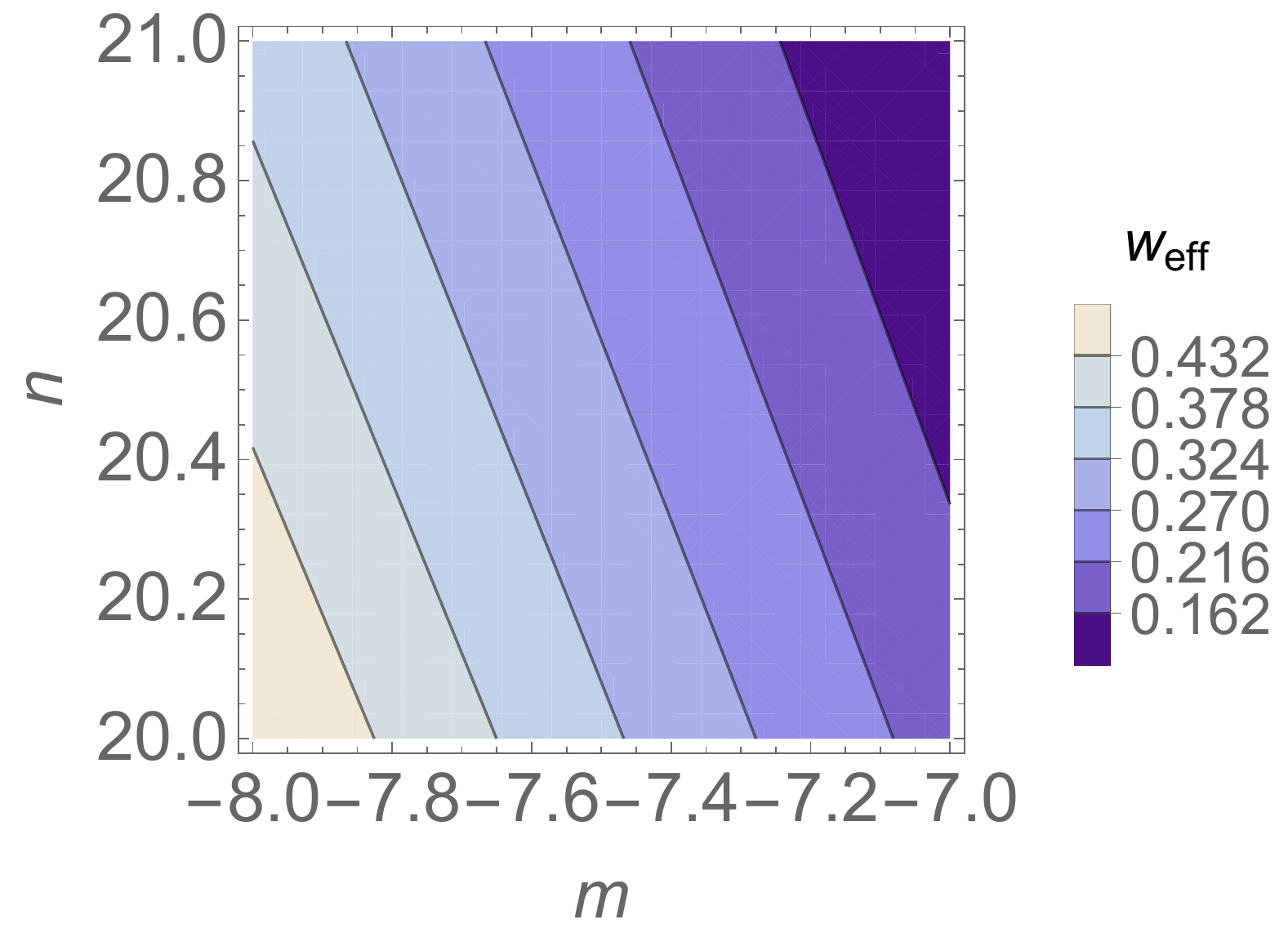}
  {}\label{fig:kl2}
\caption{The variation of the effective equation of state in a limited region in the case of $P_3$ critical point.
 }\label{fig:awage3}
\end{figure}

\par 
   \item \textit{Point $P_3$}: If we take into account the standard existence conditions, then the last critical point $P_3$ represents a solution which is characterized by a decelerated expansion of the Universe. As in the previous case, the expressions of the corresponding eigenvalues are too complex to be written here. Hence, in our analysis we rely only on numerical evaluations, determining the dynamical behavior in some limited regions. From the stability analysis, we have observed that this particular solution cannot be stable or pure unstable, and corresponds to a saddle dynamical behavior. Taking into account the existence conditions, we have displayed in Fig.~\ref{fig:awage3} the variation of the effective equation of state in this case for a limited region in the $m-n$ space. This solution corresponds to an epoch with a decelerated expansion, the evolution of the universe is highly sensitive to the values of the $m$ and $n$ parameters. Hence, this solution can explain the dust and radiation cosmological epochs in the evolution of the Universe. We have observed that for some regions the effective matter density parameter can be very close to zero and the effective equation of state can mimic a radiation behavior closely in spite of the absence of the radiation fluid.

\par
The phase space analysis at infinity is performed in the Appendix~\ref{app1}, where we show that at infinity only some of the critical points are physically viable due to the divergences of the effective equation of state.
\end{itemize}
\par
\begin{figure}[!h]
\includegraphics[height=0.4\linewidth]{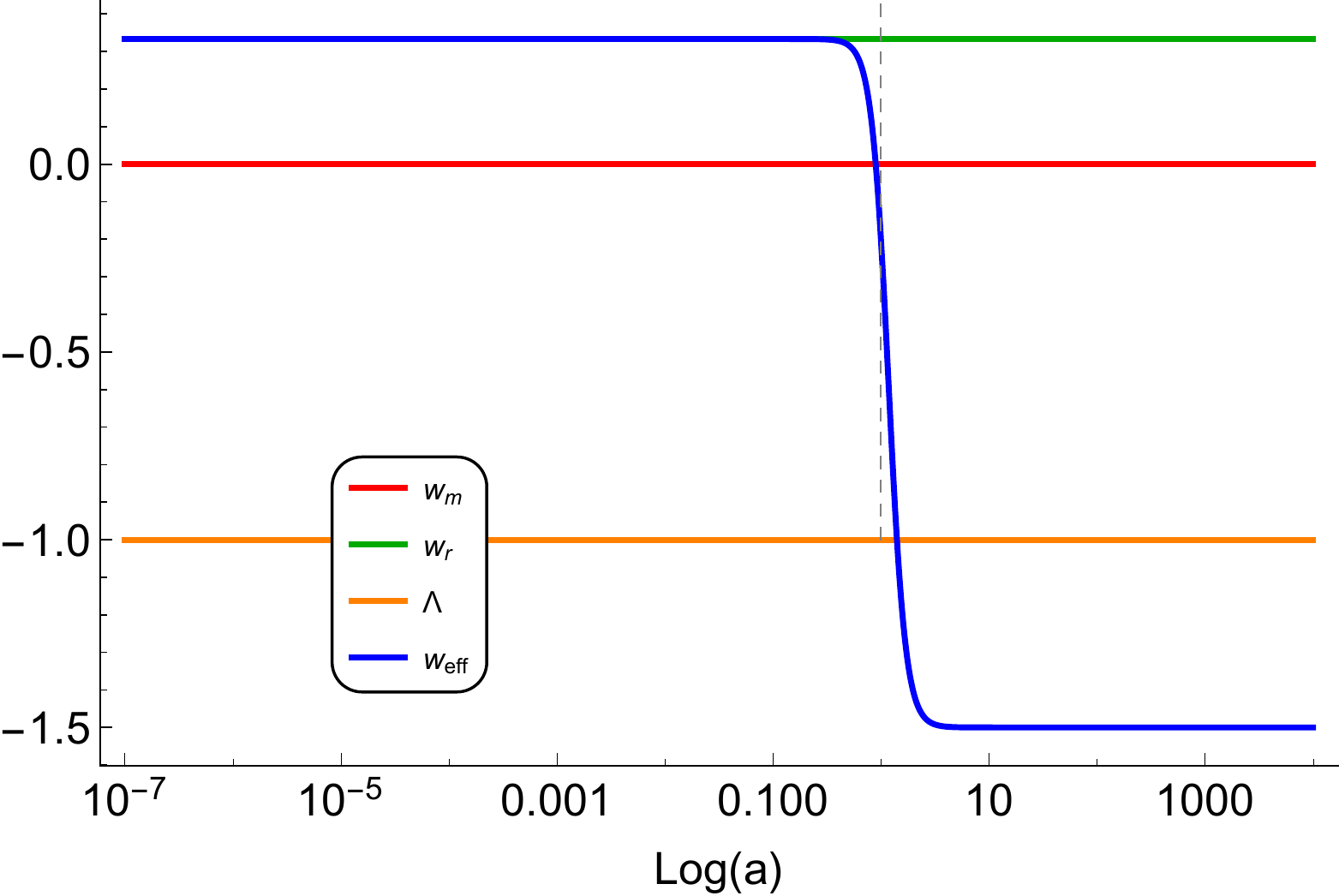}
  \caption{The evolution of the effective equation of state for the model $f(R,\mathbf{T^2})=f_0 R^{1.0001}(\mathbf{T^2})^2$.
 }\label{fig:firstmodelk}
\end{figure}

 
Finally for this model we can note that the $f(R,\mathbf{T^2})=f_0R^n(\mathbf{T^2})^m$ gravity type represents an interesting cosmological model whose viability is very sensitive to the values of the $m$ and $n$ parameters. From a dynamical analysis it can recover the accelerated expansion era and can solve the dark energy problem without introducing a cosmological constant $\Lambda$. Moreover, depending on the values of $m$ and $n$ parameters this model can recover the matter dominated epoch and act towards a stiff fluid solution. As an example, Fig.~\ref{fig:firstmodelk} shows a model for the specific case where $m=2$ and $n=1.0001$, which gives $f(R,\mathbf{T^2})=f_0 R^{1.0001}(\mathbf{T^2})^2$. This figure depicts the evolution of the effective equation of state for this model. One can notice that this case roughly describes the main epochs of our Universe, starting from a radiation dominated era with $w_{\rm eff}=1/3$, the passing to a matter dominated era with $w_{\rm eff}=0$ for a small interval, then facing an accelerating behaviour to finalizing in a super-accelerating era, the attractor of the corresponding model. As can be noted, the first model curiously exhibits the crossing of the phantom divide line boundary.

\subsection{Case 2: $f(R,\mathbf{T^2})=\alpha R^n+\beta\mathbf{(T^2)}^m$ }
\begin{figure}[!ht]
\minipage{0.5\textwidth}
  \includegraphics[height=0.8\linewidth]{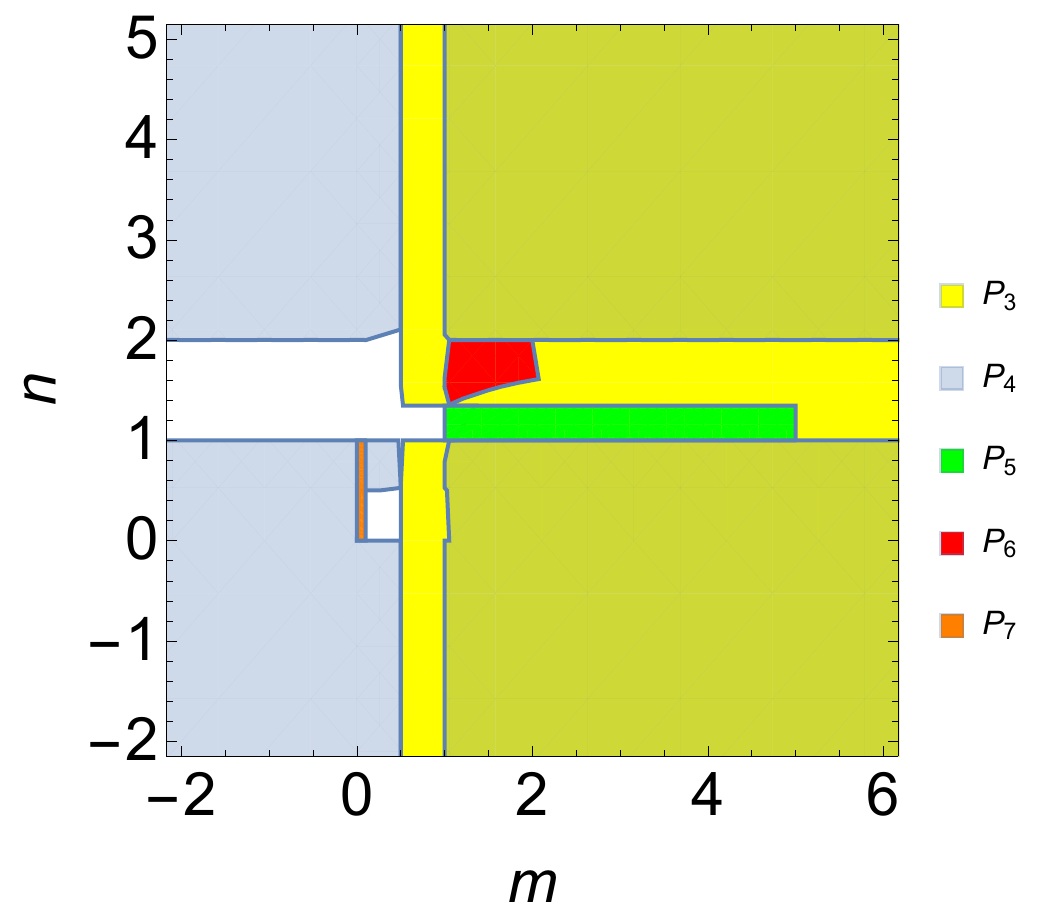}
  \caption{a)}\label{fig:st1}
\endminipage\hfill
\minipage{0.5\textwidth}
  \includegraphics[height=0.8\linewidth]{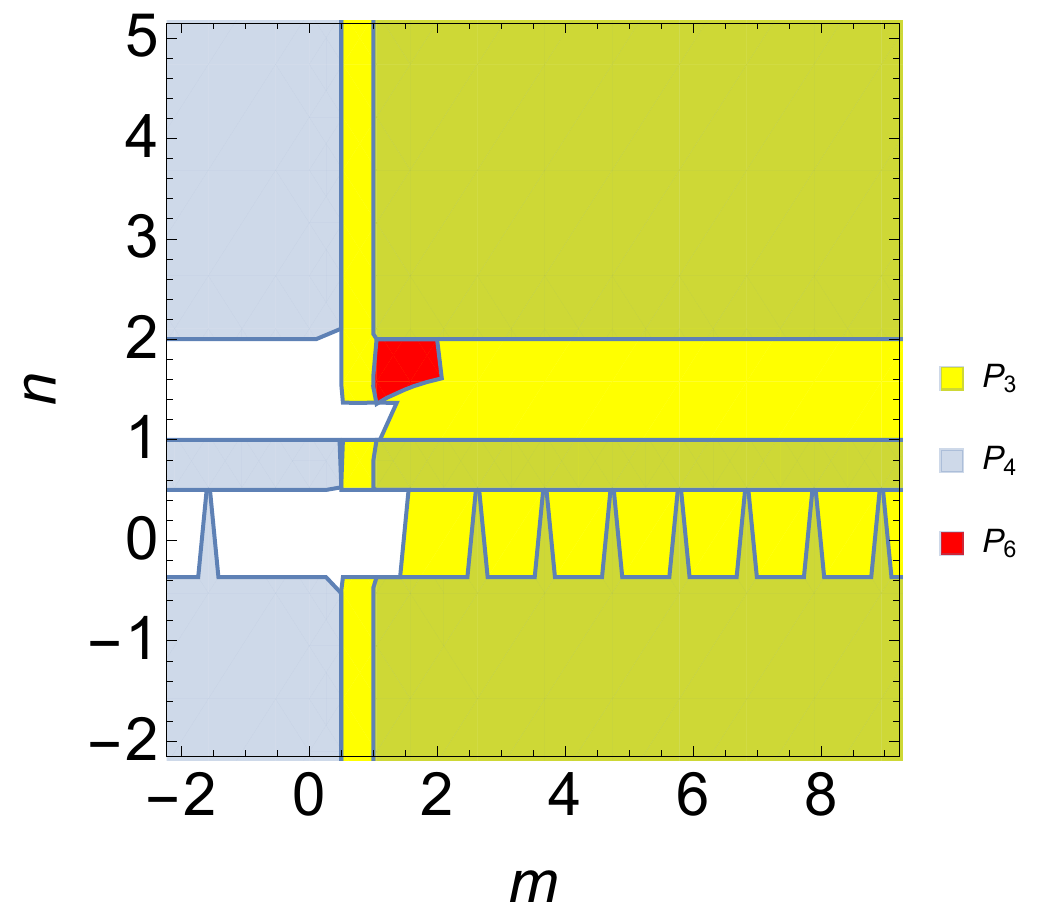}
  \caption{b)}\label{fig:st2}
\endminipage\hfill
\caption{a) Regions where the critical points are stable; b) Regions where $P_3$, $P_4$ and $P_6$ represent a stable accelerating solution}\label{fig:stability}
\end{figure}

In this section we will analyze another power-law model considering two power-law terms, the first one coming from the curvature scalar and the second one from the squared of the energy momentum tensor, implicitly given by $f(R,\mathbf{T^2})=\alpha R^n+\beta\mathbf{(T^2)}^m$
where $\alpha,\beta,m$ and $n$ are constants. We will further assume that $m\neq0$, $n\neq0$ and $n\neq 1$. For this model, it is possible to find that the dimensionless variable $x_2$ is related to the others ones, namely
\begin{equation}
 x_2=   \frac{\left(3 w^2+1\right) x_4 x_5}{2 m \left(3 w^2+4 w+1\right)}+\frac{x_3}{n}\,.
\end{equation}
Using this equation, one can straightforwardly find from \eqref{auxiliary} and \eqref{Deltais} that the auxiliary variables become
\begin{eqnarray}
   \Pi&=& -\frac{3 \left(3 w^2+4 w+1\right) x_4 \left(x_5+w+1\right)}{\left(m \left(6 w^2+8 w+2\right)-3 w^2-1\right) x_5+3 w^2+4 w+1}\,,\\
   \Delta&=&-\frac{2 (m-1) x_5 \left(x_5+w+1\right)}{x_4 \left(\left(m \left(6 w^2+8 w+2\right)-3 w^2-1\right) x_5+3 w^2+4 w+1\right)}\,,\\
   \Gamma&=&-\frac{x_4\Big(\left(4 m \left(3 w^2+4 w+1\right)+3 w^2+1\right) x_5+2 m \left(9 w^3+18 w^2+11 w+2\right)\Big)}{2 m \left(3 w^2+4 w+1\right)}-\frac{(n+1) x_3}{n}+3\,,\\
  \mho&=&\frac{1}{2} x_3\left(\frac{\left(m \left(6 w^2+8 w+2\right)-3 w^2-1\right) x_4 x_5}{m (n-1) \left(3 w^2+4 w+1\right)}+\frac{2 x_3}{n}+\frac{2 x_4}{n-1}-\frac{2}{n-1}\right)\,.
\end{eqnarray}
Hence, the dynamical system is closed and becomes a 3-dimensional one. From \eqref{dx3}--\eqref{dx5}, one gets that the final form of the dynamical system for this model becomes
\begin{eqnarray}
    \frac{dx_3}{dN}&=&x_3 \left(\frac{x_4 }{n-1}\left(\frac{\left(m \left(6 w^2+8 w+2\right)-3 w^2-1\right) x_5}{2 m \left(3 w^2+4 w+1\right)}+1\right)+\frac{x_3}{n}+\frac{1}{1-n}-2 x_3+4\right)\,,\label{dsmodelBa}\\
      \frac{dx_4}{dN}&=&x_4 \left(\frac{\left(3 w^2+1\right) x_4x_5}{2 m \left(3 w^2+4 w+1\right)}-\frac{3 \left(3 w^2+4 w+1\right) \left(x_5+w+1\right)}{\left(m \left(6 w^2+8 w+2\right)-3 w^2-1\right) x_5+3 w^2+4 w+1}+\frac{x_3}{n}-3 x_3-x_4(1+ x_5)\right)\nonumber\\
      &&+5x_4\,,\label{dsmodelBb}\\
        \frac{dx_5}{dN}&=&-\frac{3 (2 m-1) (w+1) (3 w+1) x_5 \left(x_5+w+1\right)}{\left(m \left(6 w^2+8 w+2\right)-3 w^2-1\right) x_5+3 w^2+4 w+1}\,.\label{dsmodelBc}
\end{eqnarray}
This dynamical system contains eight critical points, but only seven satisfy the condition $1\leq x_4=\Omega_{\rm m}^{\rm eff}\leq 0$. We wil now concentrate in the dust case ($w=0$). These critical points contain different types of cosmological scenarios depending on the parameters $m$ and $n$. The values of these critical points among with their effective state parameter and their acceleration conditions are displayed in Table~\ref{caseB}. In Table~\ref{caseB2} is displayed the stability criteria for each point. Let us first describe the main properties of the critical points to then analyse them one by one. The first four critical points $P_1,..,P_4$ are points which are governed by the geometrical dark energy components coming from $f(R,\mathbf{T^2})$ gravity. The origin of the phase space, the critical point $P_1$, represents a radiation era and depending on the parameters can describe a saddle point. Moreover, the critical points $P_2$ and $P_7$ also represent a radiation dominated era since their effective state parameter is equal to $1/3$. Depending on the parameters, these points can have different stability properties, but there is a family of parameters for $m$ and $n$ which ensures that all these three critical points which represent a radiation dominated era are either saddle points or unstable points. The critical points $P_3$ and $P_4$ are  purely geometric terms and exhibit similar cosmological behaviours, which can represent acceleration or not depending on the parameters. These points can represent a matter dominated era. The critical point $P_6$ always represent an accelerating scenario with a De-Sitter expansion, behaving as a cosmological constant. Each point can be summarised as follows
\begin{itemize}
    \item \textit{Point $P_1$}: The origin of the phase space exists for any values of the parameters $m$ and $n$ and always represents a radiation era. Since $w_{\rm eff}=1/3$, there is no acceleration for this critical point. This point can be either a saddle point or unstable.
     \item \textit{Point $P_2$}: This point corresponds to a universe governed only by the $\mathbf{T}^2$ term and represents a radiation era with no acceleration. Its behaviour is similar to the point $P_1$ and it also cannot be stable and always exists.
\item \textit{Point $P_3$}: This critical point represents a universe fully governed by the curvature term and depending on the parameter $n$, can describe different eras and also different stability properties. For example, it can describe an attractor accelerating cosmological solution for $m>1/2\ \land (\frac{1}{2}<n<1\lor 2 n\geq \sqrt{3}+1\lor 2 n+\sqrt{3}\leq 1)$ or can describe a saddle matter (dust) dominated era if $\left(12 n+\sqrt{73}=7\land  m<\frac{1}{2}\right)\lor \left(12 n=\sqrt{73}+7\land  m\neq \frac{1}{2}\right)$. In addition, for $\frac{1}{2}<n<1\lor n>2$ can describe a super-accelerating universe since $\dot{H}=H^2(x_3-2)>0$. These solutions are also sometimes dubbed as ``crossing the phantom divide line". Moreover, when one adds the extra condition that $m>\frac{1}{2}$, this super-accelerating solution is stable. However, this point cannot represent inflation since it is not possible to find a combination of $m,n$ such that one gets an early accelerating repeller.
\item \textit{Point $P_4$}: This critical point has similar properties as $P_3$, i.e., it can describe accelerating solutions being stable or can describe a  matter dominated era being a saddle point, or even can describe super-accelerating solutions depending on the parameters $m$ and $n$. The difference of this critical point with respect to $P_3$ is that $P_4$ does not represent a universe only governed by a curvature term since a contribution from the $\mathbf{T}^2$ is present. 
        \item \textit{Point $P_5$}: This critical point can be either stable or a saddle point and can represent a non-dust matter dominated era with $w_{\rm eff}>0$ for $0<m<\frac{1}{2}\land \frac{4 m}{2 m+1}\leq n<1$. Then, it follows that can also represent a (saddle/stable) stiff matter era when $0<m\leq \frac{1}{6}\land n=\frac{1}{2}$. It cannot represent a standard dust matter dominated era since we already assumed that $n\neq 1$. Due to the condition $0\leq\Omega_{\rm m}^{\rm eff}\leq 1$, this point cannot represent accelerating solutions either. 
         \item \textit{Point $P_6$}: This critical point represents a De-Sitter accelerating solution since $w_{\rm eff}=-1$. For different set of family of values of $(m,n)$, it can represent a late-time accelerating attractor.
          \item \textit{Point $P_7$}: This critical point corresponds to a radiation era that can be either stable or saddle depending on the parameters.
\end{itemize}
The left panel of Fig.~\ref{fig:stability} shows the regions where the critical points $P_3,..,P_7$ are stable (or spiral stable) where one can see that there is some overlapping regions where different points are stable, but there it is not possible that all of them can be stable. The right panel of Fig.~\ref{fig:stability} shows the regions where the critical points $P_3,P_4$ and $P_6$ are stable are representing accelerating solutions. Thus, this model can reproduce different cosmological eras. For example, it can describe dark energy without a cosmological constant, or a super-accelerating era, or even a stiff dominated era.
\begin{table}[H]
\centering
  \resizebox{18cm}{!}{\begin{tabular}{||c | c |c |c |c| c|c||} 
 \hline
 Cr.P. & $x_3$ & $x_4$ & $x_5$ & Existence & $w_{\rm eff}$ & Acceleration \\ [0.5ex] 
 \hline\hline
 $P_1$ & 0 & 0 & 0 & Always& $1/3$ & Never\\ [2ex]  
 \hline
 $P_2$ & 0 & 0 & $-1$& Always & $1/3$& Never\\[2ex]
 \hline
 $P_3$ & $\frac{n (4 n-5)}{2 n^2-3 n+1}$ &0 &0 & Always& $\frac{-6 n^2+7 n+1}{6 n^2-9 n+3}$&  $2 n+\sqrt{3}<1\lor \frac{1}{2}<n<1\lor 2 n>\sqrt{3}+1$\\[2ex]
 \hline
 $P_4$ & $\frac{n (4 n-5)}{2 n^2-3 n+1}$ & 0 & $-1$ & Always& $\frac{-6 n^2+7 n+1}{6 n^2-9 n+3}$&  $2 n+\sqrt{3}<1\lor \frac{1}{2}<n<1\lor 2 n>\sqrt{3}+1$\\[2ex]
 \hline
$P_5$ & $\frac{4 n-3}{2 n}$ & $\frac{-8 n^2+13 n-3}{2 n^2}$ & 0 & $\frac{13-\sqrt{73}}{16}\leq n\leq \frac{3}{10}\lor 1\leq n\leq \frac{13+\sqrt{73}}{16}$& $\frac{1}{n}-1$& Never \\[2ex]
  \hline
    $P_6$ & 2 &$-\frac{2 m (n-2)}{n}$ & $-1$ & $0\leq -\frac{2 m (n-2)}{n}\leq 1$& $-1$ & Always\\[2ex]
    \hline
     $P_7$ & 0 &$10m$ & $-1$ &$0\leq m\leq \frac{1}{10}$& $1/3$ & Never\\[2ex]
     \hline
  \end{tabular}}
 \caption{Critical points, existence condition, effective state parameter and acceleration for the $f(R,\mathbf{T^2})=\alpha R^n+\beta(\mathbf{T^2})^m$ model considering a dust matter $w=0$.} 
 \label{caseB}
 \end{table}
 \par
\begin{figure}[!h]
\includegraphics[height=0.4\linewidth]{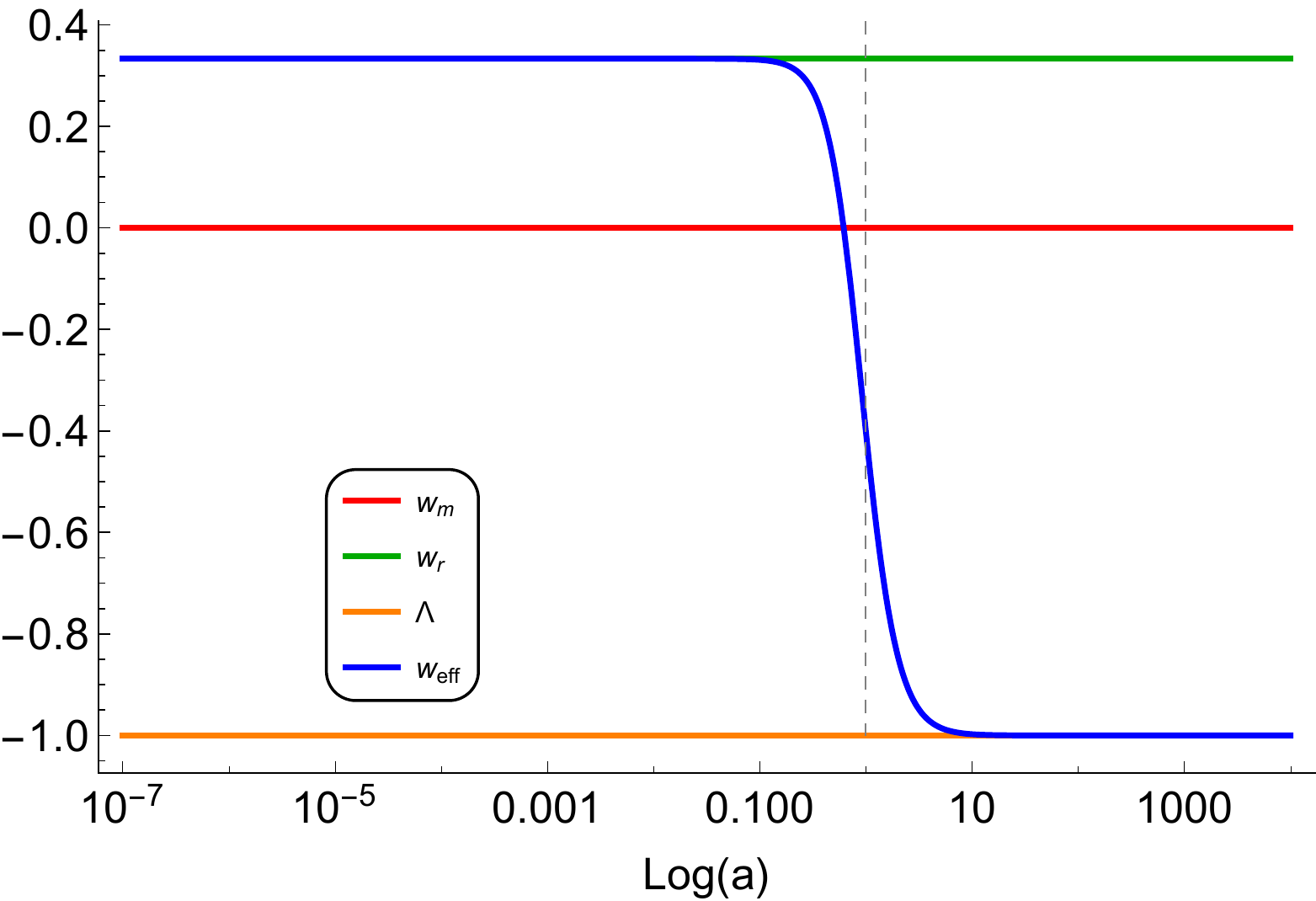}
  \caption{The evolution of the effective equation of state for the model $f(R,\mathbf{T^2})=\alpha R^2+\beta (\mathbf{T^2})^2$.
 }\label{fig:secondmodel}
\end{figure}

\begin{table}[H]
\centering
 \resizebox{18cm}{!}{\begin{tabular}{||c|c|c|c||} 
 \hline
 Cr.P. & Stable & Unstable & Saddle \\ [2ex] 
 \hline\hline
 $P_1$ &  Never & $m<1/2\land \left(n<1\lor n>5/4\right)$& $m>1/2 \lor 1<n<5/4$ \\  [2ex] 
 \hline
 $P_2$ & Never &$1/2<m<1\land \left(n<1\lor n>5/4\right)$ & $m<1/2\lor m>1\lor 1<n<5/4$   \\[2ex] 
 \hline
 \multirow{3}{*}{$P_3$} & $n<\frac{1}{16} (13-\sqrt{73})\land m>\frac{1}{2}$ & \multirow{3}{*}{$1<n<5/4\land m<1/2$} & $\left(n<\frac{1}{16} \left(13-\sqrt{73}\right)\lor \frac{1}{2}<n<1\lor n>\frac{1}{16} \left(\sqrt{73}+13\right)\right)\land m<\frac{1}{2}$ \\ [2ex] 
 & $\lor$ $\frac{1}{2}<n<1\land m>\frac{1}{2}$ &  &$\lor$ $\left(\frac{1}{16} \left(13-\sqrt{73}\right)<n<\frac{1}{2}\lor \frac{5}{4}<n<\frac{1}{16} \left(\sqrt{73}+13\right)\right)\land m<\frac{1}{2}$ \\[2ex] 
 & $\lor$ $n>\frac{1}{16} (\sqrt{73}+13)\land m>\frac{1}{2}$ &  & $\lor$ $\Big(\frac{1}{16} \left(13-\sqrt{73}\right)<n<\frac{1}{2}\lor 1<n<\frac{5}{4}\lor \frac{5}{4}<n<\frac{1}{16} \left(\sqrt{73}+13\right)\Big)\land m>\frac{1}{2}$ \\[2ex]  \hline
 \multirow{3}{*}{$P_4$} & \multirow{3}{*}{$\left(m<\frac{1}{2}\lor m>1\right)\land \left(n<0\lor \frac{1}{2}<n<1\lor n>2\right)$} & \multirow{3}{*}{$\frac{1}{2}<m<1\land 1<n<\frac{5}{4}$} & $\frac{1}{2}<m<1\land \left(n<0\lor 0<n<\frac{1}{2}\lor \frac{1}{2}<n<1\lor \frac{5}{4}<n<2\lor n>2\right)$ \\[2ex] 
 & & & $\lor \ \frac{1}{2}<m\land \left(0<n<\frac{1}{2}\lor 1<n<\frac{5}{4}\lor \frac{5}{4}<n<2\right)$ \\[2ex] 
 & & & $\lor \ m>1\land \left(0<n<\frac{1}{2}\lor 1<n<\frac{5}{4}\lor \frac{5}{4}<n<2\right)$\\[2ex] 
 \hline
  \multirow{2}{*}{$P_5$}& $m>0.5\land (1<n\lessapprox 1.327\lor 0.278\lessapprox n\leq 0.3)$ &  \multirow{2}{*}{Never} & $0.278\lessapprox n\leq 0.3\land (m<0\lor 0<m<1/2)$ \\[2ex]
 & $\lor \ m>1/2\land (1.327\lessapprox n\lessapprox1.346)$&& $\lor \ 1<n\lessapprox 1.346\land (m<0\lor 0<m<1/2)$ \\[2ex]
  \hline
    \multirow{3}{*}{$P_6$} & $0<m<\frac{1}{2}\land (1<n\leq \frac{41}{25}\lor \frac{41}{25}<n<2)$  & \multirow{3}{*}{Never}& $\left(m<-\frac{1}{2}\land 2<n\leq \frac{4 m}{2 m+1}\right)\lor \left(m=-\frac{1}{2}\land n>2\right)$   \\[2ex] 
 & $\lor \ \Big(1<m<\frac{41}{18}\land \frac{4 m}{2 m+1}\leq n\leq \frac{41}{25}\Big) \ \lor \ \Big(m=\frac{41}{18}\land n=\frac{41}{25}\Big)$ & &$\lor \ -\frac{1}{2}<m<0\land \left(n\leq \frac{4 m}{2 m+1}\lor n>2\right)$   \\[2ex] 
 & $\lor \ \Big(1<m\leq \frac{41}{18}\land \frac{41}{25}<n<2\Big)\ \lor \ \Big(m>\frac{41}{18}\land \frac{4 m}{2 m+1}\leq n<2\Big)  $& & $\lor \ \left(0<m<\frac{1}{2}\land \frac{4 m}{2 m+1}\leq n<1\right)\lor \left(\frac{1}{2}<m<1\land \frac{4 m}{2 m+1}\leq n<2\right)$ \\[2ex]
    \hline
      $P_7$ & $0<n<1\land 0\leq m\leq \frac{1}{10}$ & \multirow{2}{*}{Never}& $(n<0\lor n>1)\land 0\leq m\leq \frac{1}{10}$ \\[2ex] 
     \hline
  \end{tabular}}
 \caption{Stability of the critical points for the $f(R,\mathbf{T^2})=\alpha R^n+\beta(\mathbf{T^2})^m$ model with a pressureless matter ($w=0$). The (approximated) real numbers appearing in the inequalities above come from the numerical
analysis of the sign of the eigenvalues of the fixed point} 
 \label{caseB2}
 \end{table}
Since the phase space is non-compact for this model, one also needs to check if the critical points at infinity are physical or not. See Appendix~\ref{app2} for more details about the method used for this. One can see that there are six critical points at infinity, but four of them have a divergent $w_{\rm eff}$, hence, they are not physical. The other two critical points represent a radiation era that cannot be stable. As a numerical example, the Fig.~\ref{fig:secondmodel} shows the evolution of the effective state parameter for the model $f(R,\mathbf{T^2})=\alpha R^2+\beta (\mathbf{T^2})^2$ which corresponds to $m=n=2$. This model exhibits the three main eras of the Universe, starting form a saddle point with $w_{\rm eff}=1/3$, then a matter era with $w_{\rm eff}=0$, finalizing with a late-time de-Sitter behavior.

\section{Conclusions }

In this paper we have studied the dynamical features of a recent model of modified gravity theory which is known as the energy-momentum squared gravity model $f(R,\mathbf{T^2})$. In this model $R$ represents the scalar curvature and $\mathbf{T^2}$ the square of the energy-momentum tensor defined as $\mathbf{T^2}=T^{\mu\nu}T_{\mu\nu}$. After obtaining the field equations we then concentrated in the study of its cosmology for the standard flat FLRW spacetime. Then, we have analyzed the physical implications of such a modified gravity theory by employing the linear stability method for two specific $f(R,\mathbf{T^2})$ functions. The first scenario studied is represented by a model which takes into account a direct product between the scalar curvature $R$ and the energy momentum squared $\mathbf{T^2}$ at different real powers $n$ and $m$, with the corresponding gravity function defined as $f(R,\mathbf{T^2})=f_0 R^n(\mathbf{T^2})^m$. In this representation we have assumed that $f_0, m$ and $n$ are constant parameters. By introducing the dimensionless variables we have been able to represent the dynamics of such a modified gravity theory in flat FLRW as an autonomous system, determining the associated critical points and the corresponding stability properties. In this specific case we have observed that the phase space has a reduced complexity, with a high sensitivity to the values of the $m$ and $n$ parameters. As seen in the analysis, this specific case can recover different cosmological epochs, depending on the physical properties of the critical points obtained. 
\par 
The first critical point is a general critical point which can recover any cosmological era, staring from an accelerated expansion to matter and radiation behavior, stiff or super--stiff behavior, depending on the values of the $m, n$ constants. The second critical point is associated to an accelerated expansion, the effective matter density and total equation of state showing a high dependence on the values of the $m, n$ constants. In this case, we have determined in the figures provided specific constraints to the values of the $m, n$ constants due to the physical existence conditions and dynamical features. Finally, the third critical point represents a cosmological epoch characterized by a decelerated expansion, an era of a limited interest in the present cosmology.
\par 
The second cosmological scenario analyzed corresponds to a different mathematical model of the energy-momentum squared gravity model $f(R,\mathbf{T^2})$ which takes into account the following decomposition $f(R,\mathbf{T^2})=\alpha R^n+\beta\mathbf{(T^2)}^m$. In this case the parameters $\alpha, \beta, m$ and $n$ are assumed to be constants. This scenario shows a higher complexity of the phase space features, with various epochs corresponding to radiation, matter--dominated, de--Sitter, and solutions having accelerating or super--accelerating expansions. As noted in the manuscript, in the second cosmological scenario the values of the $m, n$ parameters dictate the phase space features and the corresponding dynamical properties associated. Furthermore, in our analysis we have obtained different numerical or relational constraints to the values of the $m$ and $n$ due to the existence conditions and physical features of the critical points corresponding to the second cosmological scenario.
\par 
The energy-momentum squared gravity theory $f(R,\mathbf{T^2})$ represents a recent proposal which takes into account the embeddedness of the energy-momentum squared scalar $\mathbf{T^2}=T^{\mu\nu}T_{\mu\nu}$ and have been studied using different approaches \cite{Board:2017ign, Moraes:2017dbs, Akarsu:2018zxl, Akarsu:2018aro,Keskin:2018bkg}. Our study is based on the linear stability method and represents a complementary analysis of the energy-momentum squared gravity theory by investigating the phase space features and the stability properties of the critical points, associated to various cosmological epochs. The analysis presented here showed that the energy-momentum squared gravity theory represents an interesting modified gravity model which can explain the current evolution of the Universe and the emergence of the accelerated expansion as a geometrical physical effect, a viable solution to the dark energy problem.

\begin{acknowledgments}
S.B. is supported by Mobilitas Pluss N$^{\circ}$ MOBJD423 by the Estonian government. P.R. acknowledges Inter University Centre for Astronomy and Astrophysics (IUCAA), Pune, India, for awarding Visiting Associateship. M.Marciu would like to thank M.C. for support and suggestions.
\end{acknowledgments}

\appendix
\section{Critical points at infinity}\label{appendix}
When a phase space is not compact, one needs to study if there are critical points at infinity. To do this, one can use the method available in~\cite{Bahamonde:2017ize,Bahamonde:2015hza,Leon:2014rra} by introducing compactified Poincaré variables. We follow this approach for the two models studied in this paper. For a 3-dimensional dynamical system with dimensionless variables $x,y$ and $z$, one can use the following Poincaré variables to compactify the phase space:
\begin{equation}
X=\frac{x}{\sqrt{1+r^2}}\,,\quad   Y=\frac{y}{\sqrt{1+r^2}}\,,\quad Z=\frac{z}{\sqrt{1+r^2}}\,,\label{coor1}
\end{equation}
where $r^2=x^2+y^2+z^2$ and then define $\rho=r/\sqrt{1+r^2}$, so that $\rho^2=X^2+Y^2+Z^2$. In these coordinates, the dynamics at infinity is recovered when $\rho\rightarrow 1$. It is then simpler to further introduce spherical coordinates
\begin{equation}
    X=\rho \cos\psi\sin\theta\,,\quad     Y=\rho \sin\theta\sin\psi\,,\quad Z=\rho \cos\theta\,, \label{coor2}
\end{equation}
where $0\leq\theta\leq 2\pi$, $0\leq\rho\leq 1$ and $0\leq \psi\leq \pi$.

\subsection{Case 1: $f(R,\mathbf{T}^2)=f_0R^n(\mathbf{T^2})^m$}\label{app1}
Following \eqref{coor1} and \eqref{coor2}, for this model one has to replace $x=x_2$, $y=x_4$ and $z=x_5$. In this case after the transformation of the dynamical equations ~\eqref{dx2}-\eqref{dx5} using the compactified Poincaré variables, we shall consider further the spherical coordinate system at infinity. Due to the high complexity of the dynamical equations before attempting to perform the limits at infinity one needs to choose specific values for the $m$ and $n$ parameters. In the case where $m=3$ and $n=2$ we obtain the following relations at infinity in the leading terms:
\begin{eqnarray}
     (1-\rho^2)\rho'&\rightarrow& 0 \,,\\
    (1-\rho^2)\theta'&\rightarrow& -\sin^2 (\theta ) \cos ^2(\theta ) \sin (\psi ) (5 \cos (2 \psi )+6)  \,,\\
   (1-\rho^2)\psi'&\rightarrow& 5 \sin (2 \theta ) \sin ^2(\psi ) \cos (\psi ) \,,
\end{eqnarray}
showing that for this specific model the angular part is decoupled. At infinity the critical points are obtained by determining the angularity of the dynamical system in the case where the right hand side of the evolution relations in the limit $\rho\rightarrow 1$ reduces to zero. In this case we have obtained the following critical points in the Poincaré variables $(X_2,X_4,X_5)$ for the general case:
\par 
\begin{equation}
    P_{\infty}^{\pm 1,2,3}=\Big\{(0,0,\pm 1),(\pm\cos\psi,\pm\sin\psi,0),(\pm\sin\theta,0,\cos\theta)\Big\}.
\end{equation}
The first two critical points at infinity, $P_{\infty}^{\pm 1}$ represent a radiation dominated epoch with a zero effective matter density parameter. The second solution at infinity,  $P_{\infty}^{\pm 2}$ does not present viable cosmological features at infinity for different values of $\psi$ due to the divergence of either the total equation of state or the effective matter density parameter. However, the last critical points $P_{\infty}^{\pm 3}$ in the case where $\theta=0$ reduces to the first critical points, $P_{\infty}^{\pm 1}$ where the geometrical dark energy component dominates and mimics a radiation era.

\subsection{Case 2: $f(R,\mathbf{T^2})=\alpha R^n+\beta\mathbf{(T^2)}^m$}\label{app2}
Following \eqref{coor1} and \eqref{coor2}, for this model one has that $x=x_3$, $y=x_4$ and $z=x_5$. By transforming the dynamical system \eqref{dsmodelBa}-\eqref{dsmodelBc} into Poincaré variables, one gets that at the limit $\rho\rightarrow 1$ (infinity), the dynamical system for the leading terms becomes
\begin{eqnarray}
    \rho'&\rightarrow&\frac{(2 m-1) \sin ^3\theta  \cos \theta \sin \psi  (n \cos (2 \psi )-n+2)}{4 m (n-1)}\,,\\
    (1-\rho^2)\theta'&\rightarrow&\frac{(2 m-1) \sin ^2\theta \cos ^2\theta  \sin \psi  (n \cos (2 \psi )-n+2)}{4 m (n-1)}\,,\\
   (1-\rho^2)\psi'&\rightarrow& \frac{(1-2 m) n \sin (2 \theta ) \sin ^2\psi  \cos\psi}{4 m (n-1)}\,,
\end{eqnarray}
where primes denote differentiation with respect to $N=\log a$. One can notice that the angular part decouples. If one sets the right hand side of these equations equal to zero, one finds three sets of critical points in Poincaré variables $(X_3,X_4,X_5)$, namely,
\begin{equation}
    P_{\infty,\pm 1}=\{0,0,\pm 1\}\,, \quad P_{\infty,\pm 2}=  \{\pm\sin\theta,0,\cos\theta\}\,,\quad P_{\infty,\pm 3}=  \{\pm\cos\psi,\pm\sin\psi,0\}\,.
\end{equation}
Since $0\leq x_4\leq 1$, one has that $0\leq (1-\rho ^2)^{-1/2}\rho  \sin\theta  \sin\psi \leq1$, and then all these critical points are in the phase space. From \eqref{weff} one can see that the effective state parameter in the Poincaré variables is
\begin{equation}
  w_{\rm eff}=  -1-\frac{2}{3}  \left(\frac{X_3}{\sqrt{1-X_3^2-X_4^2-X_5^2}}-2\right)\,.
\end{equation} This quantity is divergent for $ P_{\infty,\pm 2}$ and $ P_{\infty,\pm 3}$, hence, these critical points at infinity are non-physical unless $\theta=0$ and $\psi=\pi/2$ for $ P_{\infty,\pm 2}$ and $ P_{\infty,\pm 3}$, respectively. However, for this special choice, the Jacobian of the transformation also diverges. Then, one can conclude that only the the critical points $P_{\infty,\pm 1}$ are physical. For these critical points, the effective state parameter is equal to $1/3$ which represents a radiation era. By going back to the dynamical system for the Poincaré variables, one gets that the Eigenvalues evaluated at $P_{\infty,\pm 1}$ are
\begin{equation}
    \left\{6,\frac{16 m-11}{2 m-1},\frac{7 n-8}{n-1}\right\}\,.
\end{equation}
Then, $P_{\infty,\pm 1}$ cannot be stable, it is unstable if $\left(n<1\lor n>\frac{8}{7}\right)\land \left(m<\frac{1}{2}\lor m>\frac{11}{16}\right)$ and saddle otherwise.

\bibliography{apssamp}

\end{document}